\documentstyle[aps]{revtex} 

\def\be{\nopagebreak[3]\begin{equation}} 
\def\ee{\end{equation}} 
\def\ba{\nopagebreak[3]\begin{eqnarray}} 
\def\ea{\end{eqnarray}}

\newcommand{\teta}{\rlap{\lower2ex\hbox{$\,\tilde{}$}}\eta{}}

\newcommand{\Vec}[1]{{\mathbf{#1}}} 
\newcommand{\Vecg}[1]{{\mbox{\boldmath$ #1$}}}

\begin{document} 
\title{New Observational Bounds to Quantum Gravity Signals} 
\author{Daniel Sudarsky${}^1$ Luis Urrutia${}^1$ and H\'ector 
Vucetich${}^{2,3}$} 
\address{1. Instituto de Ciencias Nucleares\\ 
Universidad Nacional Aut\'onoma de M\'exico\\ 
A. Postal 70-543, M\'exico D.F. 04510, M\'exico\\ 
} 
\address{2. Instituto F\'\i sica\\ 
Universidad Nacional Aut\'onoma de M\'exico\\ 
A. Postal 70-543, M\'exico D.F. 04510, M\'exico\\ 
} 
\address{3. Departamento de F\'\i sica\\ 
Universidad de La Plata\\ 
La Plata, Argentina\\ 
} 
\maketitle

\begin{abstract} 
We consider a new set of effects arising from the quantum gravity 
corrections to the propagation of fields, associated with fluctuations 
of the spacetime geometry. Using already existing experimental data, 
we can put bounds on these effects that are more stringent by several 
orders of magnitude than those expected to be obtained in 
astrophysical observations. In fact these results can be already 
interpreted as questioning the whole scenario of linear (in $l_P$) 
corrections to the dispersion relations for free fields in Lorentz violating 
theories. 
\end{abstract} 





The search for experimental clues about the nature of quantum gravity
has been dismissed as unpractical\ for a long time by the simplistic
argument that such effects should appear only when the energy scales
of the interaction reaches the Planck scale, a realm far beyond our
experimental possibilities. Recently there has been a
revolutionary change in this conception originated { in
\cite{kostelecky1} and \cite{Quant Fluct} (See however
\cite{Sorkin}). The former propose a spontaneous violation of Lorentz
and CPT symmetries occurring at Planck scale, motivated by string
theory and parameterized by an extension of the standard model
including all possible Lorentz and CPT violating interactions. This
approach has sparkled a number of experimental studies of such
violations. The latter generically point out that } quantum gravity
should predict slight deviations in the laws describing the
propagation of photons in vacuum and that the cosmological distances
traveled by these gamma rays could amplify such effects making them
observable. Such modifications have been found within two currently
popular approaches to Quantum Gravity: Loop Quantum Gravity
\cite{LoopQG} and String Theory \cite{StringT}.  { These effects
predict} a change in the dispersion relations of photons that leads to
their velocity of propagation becoming energy dependent via
corrections of the type $\left( E\ell _{P}\right) ^{n}$, where $\ell
_{P}\;$is the Planck length. Observational bounds upon some of the
corresponding parameters have been settled in
Refs. \cite{BILLER,observ,KOSTELECKYO}. For example, by considering
the change in the arrival time of these gamma rays, induced by such
energy dependence, together with the intrinsic time structure of the
corresponding GRB's\cite{BILLER}, it is possible to find by a simple
order of magnitude estimate that one is bringing quantum gravity to
the realm of experimental physics!

The basic point of this letter is that if a theory predicts that
photons propagate with an energy-dependent velocity $v(E)$ rather than
with the universal speed of light $c$, this implies a breakdown of
Lorentz invariance, either fundamental or spontaneous, since such
statement can be at best valid in one specific inertial frame. This
selects a preferred frame of reference, where the particular form of
the corrected equations of motion are valid, and one should then be
able to detect the laboratory velocity with respect to that frame. It
should not be at all surprising that on the rebirth of an Ether like
concept requiring a privileged rest frame--paradoxically inspired by
current attempts to obtain a quantum description of general
relativity-- the ghost of Michelson-Morley's search should be coming
back for a revenge. Furthermore, we have today, in contrast with the
situation at the end of the 19th century, a rather unique choice for
that ``preferred inertial frame'': the frame where the Cosmic
Microwave Background (CMB) looks isotropic. Our velocity $\Vec{w}$
with respect to that frame has already been determined to be
$w/c\approx 1.23 \times 10^{-3}$ by the measurement of the dipole term
in the CMB by COBE, for instance \cite{COBE} .  From the above
discussion, it follows that the quantum gravity corrections to the
corresponding particle field theory (photons, {fermions} and others)
should contain $\Vec{w}$-dependent terms when described in our
laboratory reference frame. These would lead to a breakdown of
isotropy in the measurements carried out on earth. Thus, high
precision tests of rotational symmetry, using atomic and nuclear
systems should serve to test some of the quantum gravity corrections.
The purpose of this letter is to point out that such type of
experiments\cite{HUDR,Exper2} are sufficiently accurate to establish
bounds upon the above mentioned quantum gravity effects, due to the
very high degree of sensitivities that have been achieved. We should
point out that although the present analysis will focus specifically
in Loop Quantum Gravity inspired scenarios the same considerations
should apply {\it mutatis mutandis} to String inspired models of such
effects.

The method of analysis can be thought to correspond to the application 
of the general framework described in {the works 5 and 6 of \cite{kostelecky1}},  to the 
specific scenarios arising from the quantum gravity inspired effects. 
In the works \cite{LoopQG}, inspired on the Loop Quantum Gravity 
approach, the effects of the quantum fluctuations of the ``spacetime 
metric'' in a semiclassical state of the geometry, leave their mark on 
the effective Hamiltonian of the Maxwell field, that propagates in the 
corresponding spacetime. Such quantum gravity modifications to 
Maxwell's equations has been extended also to two-component spin 1/2 
massive particles which can be physically realized as  
neutrinos \cite{URRU1}.

Even though Maxwell theory can be considered as the paradigm for 
studying such quantum gravity corrections, it turns out that those 
which affect Dirac particles are the ones that in this case 
produce experimentally interesting effects, as we will see in the 
sequel. The starting point are the modified equations for a 
two-component\ spinor $\xi \;$with positive chirality ($\gamma 
_{5}=+1$)\ derived in \cite{URRU1}. In the nuclear case, which is of 
interest for our purposes, the scale ${\cal L\;}$ of that reference 
has the natural choice ${\cal L}$ =$1/m\;$, where $m$ is typically the 
particle mass. Also the kinetic energies involved are small compared 
to the mass so that we can safely set $\nabla ^{2}<<m^{2}.$ In this 
way the relevant equations reduce to 
\be 
\left[ i\frac{\partial }{\partial t}-i{A}\ {\Vecg{\sigma}}\cdot \nabla +\frac{%
{K}}{2}\right] \xi -m\left( \alpha -\beta i\ {\Vecg{\sigma}}\cdot \nabla 
\right) \chi =0,\qquad \left[ i\frac{\partial }{\partial t}+i{A}\ {\Vecg{%
\sigma}}\cdot \nabla -\frac{{K}}{2}\right] \chi -m\left( \alpha -\beta i\ {%
\Vecg{\sigma}}\cdot \nabla \right) \xi =0, 
\ee 
where 
\begin{equation} 
{A}=\left( 1+{\Theta_1\;}m\ell _{P}\right) ,\quad \alpha =\left( 
1+{\Theta_3\;}m\ell 
_{P}\right) ,\quad {K}=m{\Theta_4}m\ell _{P} 
,\qquad \beta =\Theta_2\,\ell 
_{P}, 
\end{equation} 
and $\Theta_1,\Theta_2,\Theta_3,\Theta_4$\ are constants. We 
are interested here in analyzing only the corrections which are linear 
in $\ell _{P}.$ In the two-component case we had $\chi =-i\sigma 
_{2}\xi ^{\ast }$. From now on we consider $\xi $\ and $\chi \;$to be 
independent spinors and we rewrite the above set of equations in terms 
of the four component spinor $\Psi ^{T}=(\xi ^{T}\;,\chi ^{T})$. This 
leads to a modified Dirac equation 
\begin{equation} 
\left( i\gamma ^{\mu }\partial _{\mu }\,+\Theta_1 m\ell 
_{P}\;i\Vecg{\gamma}\cdot 
\nabla -\frac{K}{2}\gamma _{5}\gamma ^{0}-m\left( \alpha -i\Theta_2 \ell 
_{P}\,{\Vecg{%
\Sigma}}\cdot \nabla \right) \right) \Psi =0, \label{DIREQ} 
\end{equation} 
where we have used the representation in which $\gamma _{5}$ is 
diagonal and the spin operator is\ $\Sigma ^{k}=(i/2)\epsilon 
_{klm}\gamma ^{l}\gamma ^{m}$. The normalization has been chosen so that in the limit
$(m\ell_P)\rightarrow 0$ we recover the standard massive Dirac equation. The term $m\left(
1+{\Theta_3\;}m\ell _{P}\right) $ can be interpreted 
as a renormalization of the mass whose physical value is taken to be 
$M=m\left( 1+{\Theta_3\;}m\ell _{P}\right) $. After this 
modification the effective Lagrangian is 
\begin{equation} 
L_{D}=\frac{1}{2}i\bar{\Psi}\gamma ^{0}\left( \partial _{0}\Psi \right) +%
\frac{1}{2}i\bar{\Psi}\left(\frac{}{}(1+\Theta_1M\ell _{P})\gamma^k-\Theta_2\ell _{P}M\Sigma 
^{k}\, \right)\, \partial _{k}\Psi -\frac{1}{2}M\bar{\Psi}\Psi-
\frac{K}{4}\bar{\Psi}\gamma_5\gamma^0\Psi +{\rm h.c.}. 
\label{DIRLAG} 
\end{equation} 
This Lagrangian is not Lorentz invariant and thus 
corresponds to the Lagrangian associated with time evolution as seen 
in the CMB frame. In order to obtain the Hamiltonian corresponding to 
time evolution as seen in the laboratory frame, we write 
(\ref{DIRLAG}) in a covariant looking form, by introducing explicitly 
the CMB frame's four velocity $W^{\mu }=\gamma (1,\,{\Vec{w}}/c)$. In the metric with
signature $-2$ the 
result is 
\begin{eqnarray} 
L_{D}&=&\frac{1}{2}i\bar{\Psi}\gamma ^{\mu }\partial _{\mu }\Psi -\frac{1}{2}M%
\bar{\Psi}\,\Psi +\frac{1}{2}i(\Theta_1 M\ell _{P})\bar{\Psi}\gamma _{\mu }\left( 
g^{\mu \nu }-W^{\mu }W^{\nu }\right) \partial _{\nu }\Psi \nonumber \\ 
&&+\frac{1}{4}(\Theta_2M\ell_{P})\bar{\Psi}\epsilon _{\mu \nu \alpha \beta }W ^{\mu
}\gamma ^{\nu 
}\gamma ^{\alpha }\partial ^{\beta }\Psi 
- \frac{1}{4}(\Theta_4 M\ell_P)M W_\mu \bar{\Psi}\gamma_5\gamma^\mu\Psi +h.c. . 
\label{otra} 
\end{eqnarray} 
Using the method of \cite{Lane} we obtain the non-relativistic limit of the Hamiltonian
corresponding to (\ref{otra}), to first order in $\ell_P$. To this end we make the identifications
$a_{\mu }=H_{\mu \nu }=d_{\mu \nu 
}=e_\mu=f_\mu=0,\,c_{\mu \nu }= \Theta_1 M\ell _{P}\,(g_{\mu \nu }-W_\mu W_\nu) 
$, $g_{\alpha \beta \gamma }=- \Theta_2 M\ell_{P} \,W^{\rho 
}\epsilon _{\rho \alpha \beta \gamma }$ and $b_\mu=\frac{1}{2}\Theta_4 M^2 \ell_PW_\mu$.
From Eq.(26) of \cite{Lane} we obtain, up to order ${({\bf w})/c}^{2}$, such that 
$W^{\mu }=(1+{1}/{2}\,\left({\Vec{w}}/{c} \right) 
^{2},\;{\Vec{w}}/{c})$, 
\begin{eqnarray} 
\tilde{H}&=& \left[ Mc^2(1+\Theta_1\, M\ell_P\,\left({\Vec{w}}/{c} \right) 
^{2}) 
+ \left(1+2\,\Theta_1M\ell _{P}\left(1+\frac{5}{6}\left({\Vec{w}}/{c} \right) 
^{2}\right)\right)\left(\frac{p^{2}}{2M}+g\,\mu 
\,{\bf s}%
\cdot {\bf B}\right) \right]+\nonumber\\ 
&& + \left(\Theta_2+\frac{1}{2}\Theta_4 \right)M\ell _{P}\left[\left(2Mc^2 
-\frac{2p^{2}}{3M}\right)\,{\bf s}%
\cdot \frac{\bf w}{c}+\frac{1}{M}\,{\bf s}\cdot {Q}_{P}\cdot \frac{\bf w}{c}\right] 
+\Theta_1M\ell _{P} %
\left[ \frac{{\bf w}\cdot {Q}_{P}\cdot {\bf w}}{Mc^2}\right], 
\label{B6} 
\end{eqnarray} 
where ${\bf s}={\bf \sigma}/2$. Here we have not written the terms 
linear in the momentum since they average to zero. In (\ref{B6}) $g$ 
is the standard gyromagnetic factor, and $Q_{P}$ is the momentum 
quadrupole tensor with components $Q_{Pij}=p_{i}p_{j}-1/3p^{2}\delta 
_{ij}$. The terms in the second square bracket represent a coupling of 
the spin to the velocity with respect to the ``rest'' (privileged) 
frame. The first one, originally proposed in reference 
\cite{PhilWool}, has been measured with high accuracy in references 
\cite{HUDR} where an upper bound for the coefficient has been 
found. The second term is a small anisotropy contribution and can be 
neglected. Thus we find the correction 
\be 
\delta H_{S}= \left(\Theta_2 +\frac{1}{2}\Theta_4\right) 
M\ell _{P} (2Mc^2) \left[ 1 + O\left( 
\frac{p^{2}}{2M^2c^2}\right) \right] 
\Vec{s}\cdot \frac{\Vec{w}}{c}. 
\ee 

Let us concentrate now on the last term of (\ref{B6}), which 
represents an anisotropy of the inertial mass, that has been bounded 
in Hughes-Drever like experiments. With the approximation 
$Q_{P}=-5/3<p^{2}>Q/R^{2}$ for the momentum quadrupole moment, with 
$Q$ being the electric quadrupole moment and $R$ the nuclear radius, 
we obtain 
\begin{equation} 
\delta H_{Q}=-\Theta_1 M\ell _{P}\frac{5}{3}\left\langle \frac{p^{2}}{2M}%
\right\rangle \left( \frac{Q}{R^{2}}\right) \left( \frac{w}{c}\right) 
^{2}P_{2}(\cos \theta ), 
\label{QMM} 
\end{equation} 
for the quadrupole mass perturbation, where $\theta$ is the angle 
between the quantization axis and $\Vec{w}$. Using $<p^{2}/2M>\sim 40$ 
MeV for the energy of a nucleon in the last shell of a typical heavy nucleus, 
together with the experimental bounds of references \cite{Exper2} we find 
\begin{equation} 
\mid \Theta_2+\frac{1}{2}\Theta_4\mid <2\times 10^{-9},\qquad 
\mid \Theta_1\mid <3\times 10^{-5}. \label{Result2} 
\end{equation} 
Equation (\ref{Result2}) is the main result of this 
paper. 


A second possibility to look for experiments constraining the quantum 
gravity corrections to particle interactions is provided by the 
electrodynamics of Gambini-Pullin \cite{LoopQG}. The effective Lagrangian 
density is \cite{URRU2} 
\begin{equation} 
L=\frac{1}{2}\left( E_{i}E_{i}-B_{i}B_{i}\right) -4\pi \left( \phi \rho 
-J_{k}A_{k}\right) +\theta \ell _{P}\left( E_{i}\epsilon _{ipq}\partial 
_{p}E_{q}-B_{i}\epsilon _{ijk}\partial _{j}B_{k}\right) , \label{LAGMAX} 
\end{equation} 
which is clearly not Lorenz invariant and thus must correspond to the 
Lagrangian associated with time evolution as seen in the CMB frame. We 
rewrite the Lagrangian (\ref{LAGMAX}), to order $%
\ell _{P}$, in a covariant looking form, by introducing explicitly the 
privileged frame's four velocity $W^{\mu }$, obtaining 
\begin{equation} 
L=-\frac{1}{4}F_{\mu \nu }F^{\mu \nu }-\theta \ell _{P}U^{\beta \mu \psi 
\delta \nu }\;F_{\beta \mu }\partial _{\psi }F_{\delta \nu }-4\pi J^{\mu 
}A_{\mu },\qquad U^{\tau \theta \psi \delta \nu }=\left( ^{\;}K^{\tau \theta 
\psi \delta \nu }+\frac{1}{4}\epsilon _{\;\;\;\;\;\beta \mu }^{\;\tau \theta 
\;\;}K^{\beta \mu \psi \zeta \eta }\epsilon _{\;\zeta \eta }^{\;\;\;\;\delta 
\nu }\right) , \label{EMLAG} 
\end{equation} 
with 
\begin{equation} 
K^{\beta \mu \psi \delta \nu }=\frac{1}{4}W_{\alpha }\left( \epsilon 
^{\alpha \beta \psi \delta }W^{\mu }W^{\nu }-\epsilon ^{\alpha \mu \psi 
\delta }W^{\beta }W^{\nu }+\epsilon ^{\alpha \mu \psi \nu }W^{\beta 
}W^{\delta }-\epsilon ^{\alpha \beta \psi \nu }W^{\mu }W^{\delta }\right) . 
\end{equation} 

Following the standard Noether procedure for Lagrangians depending 
upon the second derivatives of the basic field $A_{\mu }$ we can 
calculate the modified energy-momentum tensor for the Maxwell field in 
the laboratory. To first order in $\;\ell _{P}$ and to second order in 
the velocity $\Vec{w}$, the corresponding Hamiltonian density is 
\be 
T^{00}=\frac{1}{2}\left( E^{2}+B^{2}\right) -\theta\ell_P \Vec{E}\cdot 
\nabla \times \Vec{E}+\theta \ell_P \Vec{B}\cdot \nabla \times 
\Vec{B}- \frac{2 \theta \ell_P }{c}\frac{\Vec{w}}{c}\cdot 
\Vec{E}\times \frac{\partial \Vec{E} }{\partial 
t}+\frac{w}{c}^{i}\ell_P\,R_{ij}(E,B,\partial _{k})\frac{w}{c}^{j}+O(%
\frac{w}{c}^{3}). 
\label{T00EM} 
\ee In principle it seems worthwhile to consider the effect of time
dependent fields, such as the fourth term above, in experiments
designed to test the isotropy of the laws of physics. As far as we
know, no experiment has been performed up to this date that could
detect such effects, and so it would be very interesting to analyze
the degree to which, these predictions can be tested with the current
available technology. The quadratic piece $R_{ij}$ in (\ref{T00EM})
includes only parity-violating terms which do not produce additional
contributions to the quadrupolar mass modifications (\ref{QMM}). Our
corrections (\ref{EMLAG}) to the Maxwell action are power counting
non-renormalizable and { being considered to be highly suppressed
in the standard model extension of Ref.\cite{kostelecky1}, they are
left out in the works \cite{electro}. Recently such terms
have been considered in \cite{KOSTELECKY} in relation to the problems
of stability and microcausality of the theory at high energies}. The
term proportional to $g_{\alpha\beta\gamma}$ in (\ref{otra}) is {also} excluded from the standard model extension,  { because it is
incompatible with the electroweak structure, as stated in the work
6 of \cite{kostelecky1}}. However, it has been subsequently considered
in Ref.\cite{KOSTELECKYO} for the case of protons and neutrons,
because of the composite nature of these particles. From our
perspective this term should always be present and in our analysis it
is responsible for the correction $\delta H_{S}$, leading to one of
the bounds established in this paper.

We have found that after identifying the preferred frame of reference 
associated with Planck scale physics effects with the CMB frame, 
existing results of atomic and nuclear physics experiments can be 
translated into very strict bounds on the quantum gravity induced 
modifications to the propagation of Dirac fields. This is a 
remarkable case in which the interplay of cosmology, atomic and 
nuclear physics serves to shed light on a field that is usually 
considered to be beyond the realm of experimental physics, namely 
quantum gravity. Moreover, the resulting bounds of order $10^{-5} $ 
and $10^{-9}$ on terms that were formerly expected to be of order 
unity, already call into question the scenarios inspired on the 
various approaches to quantum gravity, suggesting the existence of 
Lorentz violating Lagrangian corrections which are linear in Planck's 
length. 
{ This would not apply, however, to $\ell_P$-dependent Lorentz
covariant theories \cite{Smolin}.}
Alternatively, we could view the existence of such bounds on the 
linear 
corrections as wanting for an explanation for the appearance of yet 
one more unnaturally small number in the fundamental laws of physics. 

\section*{Acknowledgments}

D.S. would like to acknowledge partial support from DGAPA--UNAM Project No 
IN 112401 and CONACYT project 32272-E. L.U acknowledges partial support from 
DGAPA--UNAM Project No IN-117000 and CONACYT project 32431-E. We thank J. 
Hirsch for useful conversations and R. Lehnert for pointing out some 
problems in an 
earlier version, together with relevant suggestions.

\end{document}